\documentstyle[12pt]{article}

\advance\hoffset by -1.5cm
\def\3{\ss}
\def\sq{\hbox{\rlap{$\sqcap$}$\sqcup$}}
\def\qed{\ifmmode\sq\else{\unskip\nobreak\hfil
\penalty50\hskip1em\null\nobreak\hfil\sq
\parfillskip=0pt\finalhyphendemerits=0\endgraf}\fi}
\def\half {\frac{1}{2}}
\def\thalf {\frac{3}{2}}

\def\bbbr {{\rm I\!R}}
\def\bbbn {{\rm I\!N}}
\def\bbbone {{\mathchoice {\rm 1\mskip-4mu l} {\rm 1\mskip-4mu l}
{\rm 1\mskip-4.5mu l} {\rm 1\mskip-5mu l}}}
\def\bbbc{{\mathchoice {\setbox0=\hbox{$\displaystyle\rm C$}\hbox{\hbox
to0pt{\kern0.4\wd0\vrule height0.9\ht0\hss}\box0}}
{\setbox0=\hbox{$\textstyle\rm C$}\hbox{\hbox to0pt{\kern0.4\wd0\vrule
height0.9\ht0\hss}\box0}} {\setbox0=\hbox{$\scriptstyle\rm
C$}\hbox{\hbox to0pt{\kern0.4\wd0\vrule height0.9\ht0\hss}\box0}}
{\setbox0=\hbox{$\scriptscriptstyle\rm C$}\hbox{\hbox
to0pt{\kern0.4\wd0\vrule height0.9\ht0\hss}\box0}}}}

\newtheorem{Theorem}{Theorem}

\begin{document}
\thispagestyle{empty}
\begin{flushright}
DAMTP-94-51 \\
hep-th/9407186
\end{flushright}
\vspace{2.0cm}

\begin{center}

{\Large An explicit construction of the quantum
group in chiral WZW-models}
\vspace{2.0cm}

{\large Matthias R. Gaberdiel}
\footnote{e-mail: M.R.Gaberdiel@amtp.cam.ac.uk} \\
{Department of Applied Mathematics and Theoretical
Physics\\
University of Cambridge, Silver Street \\
Cambridge, CB3 9EW, U.\ K.\ }
\vspace{0.5cm}

{\bf Abstract}
\end{center}

It is shown how a chiral Wess-Zumino-Witten theory
with globally defined vertex operators and a one-to-one
correspondence between fields and states can be constructed.
The Hilbert space of this theory is the direct sum of tensor products
of representations of the chiral algebra and finite dimensional
internal parameter spaces. On this enlarged space there exists
a natural action of Drinfeld's quasi quantum group $A_{g,t}$, which
commutes with the action of the chiral algebra and plays the r\^{o}le
of an internal symmetry algebra. The $R$ matrix describes the braiding
of the chiral vertex operators and the coassociator $\Phi$ gives rise
to a modification of the duality property.

For generic $q$ the quasi quantum group is isomorphic to the
coassociative quantum group $U_{q}(g)$ and thus the duality property
of the chiral theory can be restored. This construction has to be
modified for the physically relevant case of integer level.
The quantum group has to be replaced by the corresponding
truncated quasi quantum group, which is not coassociative because of
the truncation. This exhibits the truncated quantum group as the
internal symmetry algebra of the chiral WZW model, which therefore has
only a modified duality property. The case of $g=su(2)$
is worked out in detail.

\section{Introduction}

A very important feature of two-dimensional conformal field theory is
the fact that the theory ``factorises'' into a holomorphic and an
anti-holomorphic theory. These two subtheories correspond essentially
to the left- and right-movers of the original classical theory and are
analytic (anti-analytic) in the sense that all correlation functions
are meromorphic functions of the analytic (anti-analytic) parameters.
Many properties of conformal field theory can be studied separately
for the two chiral theories.  This is of great importance as it allows
the use of the powerful methods of complex analysis for the analysis
of conformal field theory.

However, the process of breaking up a theory into the two chiral
theories is not very well understood. In particular, the naive chiral
theory, i.~e.\ the theory, in which the Hilbert space is just the
direct sum of the chiral representation spaces, does not possess
globally defined vertex operators or a one-to-one correspondence
between vertex operators and states.
\medskip

In this paper I want to show how to construct a chiral theory with a
proper Hilbert space formulation for a WZW conformal field theory. The
basic idea is to use the Hilbert space formulation of the whole theory
and to restrict it to a chiral subtheory by taking a suitable limit.
This construction guarantees that the chiral theory has a one-to-one
correspondence of states and vertex operators and that the chiral vertex
operators are well-defined operators on the whole chiral Hilbert space.
This Hilbert space is larger than the ``naive chiral Hilbert space'',
i.~e.\ the direct sum of the irreducible representations of the chiral
algebra, and is precisely the Hilbert space Moore and Reshetikhin
\cite{MR} postulated some years ago.  The additional degrees of
freedom keep track of the different ``chiral vertex operators'' of
Moore and Seiberg \cite{MS} and make sure that the vertex operators
are well-defined on the whole Hilbert space.  They furthermore retain
sufficient information to reconstruct the whole theory from its chiral
subtheory.
\smallskip

On the additional degrees of freedom there is a natural action of the
quasi quantum group $A_{g,t}$ of Drinfel'd \cite{D1}, which commutes
with the chiral algebra.  Chiral vertex operators transform
covariantly under the quasi quantum group and the braiding of the
chiral vertex operators is described by the $R$-matrix, where the
deformation parameter $h$ is related to the level of the affine
algebra by $h=2 \pi i / (k + h^{*})$ with $h^{*}$ the dual Coxeter
number of $g$.

The symmetry algebra is only a {\em quasi} Hopf algebra. This means
that the algebra is not coassociative, but only coassociative up to
conjugation. The operator, by which the two different actions on a
triple tensor product have to be conjugated, is called
the coassociator $\Phi$. It is an invertible element
in the triple tensor product of the quasi Hopf algebra.
The property of the quasi quantum group to be only quasi coassociative
leads to a modification of the duality property of the chiral
vertex operators: the two different ways of writing the operator
product are related by the action of $\Phi$ on the internal degrees of
freedom.
\smallskip

For generic $q$ the quasi Hopf algebra is
isomorphic to the (coassociative) quantum group $U_{q}(g)$. I can thus
use this isomorphism to regard the internal parameter
spaces as representations spaces of this quantum group. The chiral
vertex operators then satisfy the (unmodified) duality property
and the chiral theory transforms
naturally under the action of the quantum group.

The above construction can be extended to the physically relevant
case, where $k$ is an integer and $q$ a root of unity.
In this case one has to replace the quantum group $U_{q}(g)$
by the corresponding truncated quantum group. This is the so-called
``weak quasi-triangular quasi Hopf algebra'' canonically associated to
$U_{q}(g)$ and has been studied in \cite{MaS1}.
It is not a coassociative algebra, as the truncation breaks the
coassociativity. This exhibits the truncated quantum group as
the internal symmetry algebra of the chiral theory, which thus
possesses only a modified duality property.

On the other hand, in contrast to the original quasi quantum
group $A_{g,t}$, the degree by which the truncated quantum group
fails to be coassociative can be easily determined.
In particular, the $\Phi$-map, which
describes the non-coassociativity, is trivial on all
triple tensor products, which do not exhibit any truncation.
\medskip

I would like to mention that the general structure has been
conjectured among others by \cite{MR,AGS1,AGS2} --- for a historical
review see for example \cite{GS3}. There have been attempts to give a
construction of the quantum group in conformal field theory in
\cite{GS1,GS2} for the minimal models and in \cite{RRR1,RRR2} for the
WZW-models using the Coulomb gas picture.  However, in these
formulations the quantum group generators do not commute with the
Virasoro algebra and
the construction relies on a specific realisation of the theory (which
only exists for generic $q$).

The construction of the chiral theory (with internal degrees of
freedom) was inspired by the recent work of Chu and Goddard
\cite{MP1}, in which chiral vertex operators for $su(n)$ at
level $1$ were constructed. However, applying
my construction to the free field realisation of the whole theory
(for $su(n)$ at level $1$) does {\em not} reproduce their chiral theory.
Indeed, the internal degrees of freedom in my construction are
finite dimensional, whereas their internal parameter spaces are infinite
dimensional. On the other hand, their construction has the virtue of
preserving the (unmodified) duality property for the chiral theory.

It is a priori rather surprising that there should be two different
chiral theories, as the quantisation of a classical theory should be
somehow unique. On the other hand, this uniqueness property only
applies to the whole theory and one would expect that the
corresponding reconstructed whole theories agree. Thus the two chiral
theories just appear to be two different factorisations of the same
whole theory.

It is nevertheless quite remarkable that there exists a chiral theory
with the unmodified duality property for the level $1$ $su(n)$-theories.
However, for non-integer quantum dimensions this is only possible if
the internal degrees of freedom are infinite-dimensional. Thus chiral
theories with finite dimensional internal degrees of freedom possess
in general only a modified duality property.
\smallskip

Finally, I would like to mention that there have been attempts to
reconstruct the abstract internal symmetry algebra of a chiral
conformal field theory from the structural data of the chiral theory
\cite{Ma,V,FGV,Scho,Nill}, generalising the approach of Doplicher and
Roberts \cite{DR} to the case of braid group statistics.  This
approach has been successful in the sense that for every chiral
conformal field theory a weak quasi quantum group can be found whose
representation theory reproduces the fusion structure of the chiral
theory. However, it has been realised recently \cite{Kr,Scho} that the
abstract internal symmetry algebra is not yet uniquely determined by this
condition alone, as there exists an infinite choice for the
dimensionalities of the representation spaces.

In the above construction of the chiral WZW-model, one of the possible
abstract internal symmetry algebras is naturally selected. In addition
to satisfying all conditions of an abstract internal symmetry
algebra, it is also a subalgebra of the symmetry algebra of the whole
theory. The construction therefore seems to suggest that the internal
symmetry algebra of a chiral conformal field theory might be uniquely
determined by the whole conformal field theory, namely by the
condition to be a subalgebra of the whole symmetry algebra. The
different possible abstract internal symmetry algebras would then also
have a natural interpretation: they would correspond to different
whole conformal field theories which possess the same chiral half.
\bigskip

The paper is organised as follows. In section~2, the chiral theory is
defined. Section~3 contains a brief review of the quasi Hopf algebra
of Drinfel'd, whose appearance in the chiral theory is proven in
section~4. In the following section I explain how the duality property of
the chiral vertex operators can be restored for generic $k$
by transforming to the quantum group $U_{q}(g)$. This is demonstrated
explicitly for the case of $g=su(2)$. I then explain how the
construction has to be modified for roots of unity.
In section~6, I show how the original theory can be
reconstructed from the chiral theory by some sort of gauging
procedure. Section~7 contains some conclusions.

\section{Construction of chiral vertex operators}
\renewcommand{\theequation}{2.\arabic{equation}}
\setcounter{equation}{0}

Let us start by establishing some notation. The physical Hilbert
space of the whole (non-chiral) WZW-theory
is the finite direct sum
\begin{equation}
{\cal H} =
\bigoplus_{l} {\cal H}_{l} \otimes \overline{\cal H}_{l},
\end{equation}
where ${\cal H}_{l}$ ($\overline{\cal H}_{l}$) are irreducible (not
necessarily equivalent) representations of the chiral algebra
${\cal A}$ ($\overline{\cal A}$). The two chiral algebras ${\cal A}$
and $\overline{\cal A}$ commute and are both isomorphic to
the affine algebra $\hat{g}$, generated by
\begin{equation}
[J^{a}_{m}, J^{b}_{n}]= i\; f^{ab}_{c} J^{c}_{m+n} +
\half k\; m \; \delta_{m,-n} \delta^{ab}.
\end{equation}
$k$ ($\overline{k}$) lies in the center of the algebra
${\cal A}$ ($\overline{\cal A}$) and both $k$ and $\overline{k}$
take the same value in all representations.
The value of $x= k/ \psi^{2} $ is called the level,
where $\psi$ is a long root of the Lie algebra $g$. It has to be an
integer in a unitary theory. This restriction can also be understood
as a quantisation condition, if the WZW model is to be regarded as the
quantisation of a classical field theory (see (\ref{actionS}) below).
\smallskip

I want to assume that there is a one-to-one correspondence between
fields and states (for the whole theory) and denote for a given
$\psi\otimes \overline{\psi}\in{\cal H}_{j}\otimes \overline{\cal H}_{j}$ the
corresponding field by $\phi(\psi\otimes\overline{\psi}; z,\bar{z}): {\cal
H}\rightarrow {\cal H}$.  I also assume, that if one specifies the
``source'' and the ``range'' of the operator --- i.~e.\ if one
considers only its projected parts --- the operator
$\phi(\psi\otimes\overline{\psi}; z,\bar{z})$ decouples
into a sum of tensor products of operators,
each of which depends only on $z$ and $\bar{z}$, respectively, i.\ e.\
\begin{equation}
\label{decoup}
\left[ \phi(\psi\otimes\overline{\psi}; z,\bar{z})\;
\left( |\chi\rangle_{l} \otimes
|\overline{\chi}\rangle_{l}\right) \right] _{m} =
\sum_{\alpha} \Bigl( V^{\alpha}_{m l}(\psi,z)\; |\chi\rangle_{l}
\Bigr) \otimes \left(
\bar{V}^{\alpha}_{m l} (\overline{\psi},\bar{z}) \;
|\overline{\chi}\rangle_{l}\right).
\end{equation}
As an aside I would like to point out that (\ref{decoup}) does not
determine the operators $V^{\alpha}_{m l}(\psi,z)$ and
$\bar{V}^{\alpha}_{m l} (\overline{\psi},\bar{z})$ uniquely,
as there is (at least) the freedom to redefine for each $\alpha$
\begin{equation}
\label{redef}
\tilde{V}^{\alpha}_{m l}(\psi,z):= \lambda V^{\alpha}_{m l}(\psi,z)
\hspace{1.0cm} \widetilde{\bar{V}^{\alpha}}_{m l}
(\overline{\psi},\bar{z}) : = \lambda^{-1}
\bar{V}^{\alpha}_{m l} (\overline{\psi},\bar{z}),
\end{equation}
where $\lambda$ is a non-zero complex number.

In general $\phi(\psi\otimes\overline{\psi})$
does {\em not} decompose into such a sum
of products and thus the whole theory is not simply the product of the
holomorphic and the anti-holomorphic theory. However,
a chiral theory with a proper Hilbert space
formulation can be defined, if one enlarges the (naive chiral) Hilbert
space. The internal parameter spaces which are introduced in
this way, keep track of the different ``chiral vertex operators'' of
\cite{MS} and retain all the necessary information to reconstruct the
original theory.
\smallskip

Each irreducible representation $\overline{\cal H}_{j}$ of
$\overline{\cal A}=\hat{g}$ contains a subspace $\overline{W}_{j}$ of
highest weight vectors (w.~r.~t.\ the Virasoro algebra), which forms
a finite dimensional (irreducible) representation of the Lie algebra
$g$. This subspace can be interpreted as the space of lowest energy
states, as $\overline{L}_{0}$ is essentially the energy.

The basic idea of the construction is to enlarge the naive chiral
Hilbert space by these lowest energy states, i.~e.\ to define the
Hilbert space of the holomorphic chiral theory as
\begin{equation}
{\cal H}_{chir} = \bigoplus_{l} {\cal H}_{l}\otimes \overline{W}_{l} .
\end{equation}
On this chiral Hilbert space one can then define
globally well-defined chiral vertex operators.
These chiral vertex operators are in
one-to-one correspondence with vectors in ${\cal H}_{chir}$ and thus
depend on $\psi\otimes\bar{w} \in {\cal H}_{j}\otimes \overline{W}_{j}$.
They can be defined as
\begin{equation}
\label{chir}
V(\psi\otimes\bar{w},z): {\cal H}_{chir} \rightarrow {\cal H}_{chir}
\end{equation}
by setting
\begin{equation}
\Bigl( V(\psi\otimes\bar{w},z)\;\left( |\chi\rangle_{l} \otimes
|\bar{u}\rangle_{l}\right) \Bigr)_{m} : =
\lim_{\bar{z}\rightarrow 0}
\bar{z}^{\bar{\Delta}_{j} + \bar{\Delta}_{l} -
\bar{\Delta}_{m}}\;
\Bigl( \phi(\psi\otimes\bar{w};z,\bar{z})\;
\left( |\chi\rangle_{l} \otimes
|\bar{u}\rangle_{l} \right) \Bigr)_{m},
\end{equation}
where $\bar{\Delta}_{i}$ is the $\bar{L}_{0}$-eigenvalue of the
highest weight representation $\overline{W}_{i}$ of $\overline{\cal
H}_{i}$. A similar construction can be performed for the
anti-holomorphic vertex operators.
\smallskip

To prove, that the above definition makes sense, I have to show that
\begin{equation}
\label{claim}
\lim_{\bar{z}\rightarrow 0}
\bar{z}^{\bar{\Delta}_{j} + \bar{\Delta}_{l} -
\bar{\Delta}_{m}} \;
\bar{V}^{\alpha}_{m l}(\bar{w},\bar{z})\;
|\bar{u}\rangle_{l} \in \overline{W}_{m}.
\end{equation}
Because of the conformal invariance of the theory, the operator
product expansion is given by \cite{Peter89}
\begin{equation}
\bar{V}^{\alpha}_{m l}(\bar{w},\bar{z})\;
|\bar{u}\rangle_{l} = \sum_{r\in\bbbn_{0}} \bar{z}^{r+\bar{\Delta}_{m}
- \bar{\Delta}_{j} - \bar{\Delta}_{l}} |\bar{\chi}_{r}\rangle,
\end{equation}
where $|\bar{\chi}_{r}\rangle\in \overline{\cal H}_{m}$ has conformal
weight $\bar{\Delta}_{m}+r$.  However, this already implies that
(\ref{claim}) is satisfied, as I project onto the lowest energy states
by taking the limit $\bar{z}\rightarrow 0$.
\smallskip

This definition determines well-defined associative
operators on ${\cal H}_{chir}$. These operators are in one-to-one
correspondence with states in ${\cal H}_{chir}$.
They also satisfy certain braid relations and possess a
(modified) duality property, as shall be shown in section~4.
\medskip

It cannot surprise that the properly defined chiral vertex operators
depend on these internal parameter spaces. To see this, recall that the
WZW-model can be understood as the quantisation of the classical field
theory with action
\begin{equation}
\label{actionS}
{\cal S}[g] = k \left[ - \frac{1}{16 \pi} \int_{\cal M} tr \left(
g^{-1} \partial_{\mu} \; g^{-1} \partial^{\mu} g \right) d^{2} x +
\frac{1}{24 \pi} \int_{\cal B}
\varepsilon^{\lambda \mu \nu} tr\left(
g^{-1} \partial_{\lambda} g \; g^{-1} \partial_{\mu} g \; g^{-1}
\partial_{\nu} g \right) d^{3}x \right],
\end{equation}
where $g:{\cal M} \rightarrow G$ and ${\cal M}= S^{1} \times \bbbr$,
${\cal B}$ is a three-dimensional manifold with $\partial {\cal
B}={\cal M}$ \cite{MPHOS}.  Writing $x^{\pm} = t \pm x$, the equations
of motion are
\begin{equation}
\partial_{-} \left(\partial_{+} g g^{-1} \right) =
\partial_{+} \left(g^{-1} \partial_{-} g \right) = 0,
\end{equation}
which implies that $g(x,t)$ factors into a left- and right-moving part
\begin{equation}
g(x,t) = u(x^{+})\; v(x^{-}).
\end{equation}
However, this decomposition is only determined up to a constant group
element, as one can redefine
\begin{equation}
u \rightarrow u h \hspace{1.5cm} v \rightarrow h^{-1} v,
\end{equation}
where $h\in G$ is arbitrary.  Thus, in order to be able to reconstruct
the whole theory from its chiral components, one has to keep track of
this ambiguity when defining the chiral theory.  Quasi-classically,
one therefore expects that one has to retain internal parameter spaces
in the chiral theory, which are representations of the Lie algebra
$g$. The lowest energy vectors form indeed such spaces and the tensor
product of two such representations is just the usual $g$-tensor
product, as the comultiplication of the affine algebra \cite{MG1},
restricted to the horizontal Lie algebra, is trivial.

In the (proper) quantum theory one retains the same degrees of
freedom. However, these spaces become ``quantised'', as they should
now be regarded as representation spaces of the quasitriangular
quasi-Hopf algebra of Drinfel'd $A_{g,t}$ \cite{D1}.  This is due to
the fact that the vertex operators do not commute for finite $k$, but
only satisfy braid relations. I shall explain this in more detail in
section~4, after I have recalled the definition of this quasi quantum
group.  For the moment I only want to point out that this quasi Hopf
algebra is (up to twisting) the unique quantisation of the universal
enveloping algebra $U(g)$ \cite{D1} and that its appearance here is a
typical quantum effect, as only the lowest energy states are involved.

\section{The quasitriangular quasi-Hopf algebra $A_{g,t}$}
\renewcommand{\theequation}{3.\arabic{equation}}
\setcounter{equation}{0}

As an algebra, $A_{g,t}$ is the universal enveloping algebra $U(g)$ of
$g$. The comultiplication is defined to be
\begin{equation}
\label{com}
\Delta(a)=a\otimes\bbbone + \bbbone\otimes a ,
\end{equation}
the counit $\varepsilon(a)=0,\;\varepsilon(\bbbone)=1$ and the
antipode
\begin{equation}
\label{antip}
S(a) = -a.
\end{equation}
The $R-$matrix is given by
\begin{equation}
\label{rm}
R=\Delta(q^{C})\;\left( q^{-C}\otimes q^{-C} \right) = e^{\frac{h
t}{2}},
\end{equation}
where $C$ is the quadratic Casimir of $g$
\begin{equation}
\label{Casimir}
C=\sum_{a} \tau^{a}\;\tau^{a},
\end{equation}
and $t$ is twice the split Casimir, namely
\begin{equation}
t = \Delta(C) - C \otimes\bbbone - \bbbone\otimes C ,
\end{equation}
which satisfies $m(t)= 2 C$, where $m: U(g)\times U(g) \rightarrow
U(g)$ is the multiplication map. $q$ is a complex number and will turn
out to be
\begin{equation}
q= e^{\frac{h}{2}} = e^{\frac{\pi i}{k+h^{*}}},
\end{equation}
i.~e.\ $h=\frac{2 \pi i}{k+ h^{*}}$, where $h^{*}$ is the dual Coxeter
number of $g$. I have chosen the normalisation $Tr\; \tau^{a} \tau^{b}
= - \half \delta^{ab}$ and hence the Casimir (\ref{Casimir}) is half
the Casimir operator of Drinfel'd \cite{D2}.

$A_{g,t}$ is not a quantum group, but only a quasi-quantum group.  By
this one means, that there exists an invertible element $\Phi\in
U(g)\otimes U(g)\otimes U(g)$, $\Phi\neq
\bbbone\otimes\bbbone\otimes\bbbone$, such that
\begin{equation}
\left( id \otimes \Delta \right) \circ \Delta (a) =
\Phi \left(\Delta\otimes id \right) \circ \Delta (a) \;
\Phi^{-1}
\end{equation}
for all $a\in U(g)$. At first sight one might think, that $A_{g,t}$ is
in fact a quantum group, as the comultiplication is obviously
coassociative and thus one might choose $\Phi=\bbbone$. However, the
above $R-$matrix (\ref{rm}) does not satisfy the consistency condition
necessary for an ordinary quantum group, namely
\begin{equation}
\label{rmc}
(\Delta\otimes id)\; (R) = R_{13} \; R_{23} \hspace{1cm} (id \otimes
\Delta)\; (R)= R_{13} \; R_{12},
\end{equation}
but only the weaker condition
\begin{eqnarray}
\label{rc1}
{\displaystyle (\Delta\otimes id)\; (R)} & = & {\displaystyle
\Phi_{312} \; R_{13} \; (\Phi_{132})^{-1}\; R_{23} \; \Phi} \\
\label{rc2}
{\displaystyle (id \otimes \Delta)\; (R)} & = & {\displaystyle
(\Phi_{231})^{-1} R_{13} \; \Phi_{213} \; R_{12} \; (\Phi)^{-1} .}
\end{eqnarray}

In order for the above to be a well-defined quasitriangular quasi-Hopf
algebra, these maps must in addition satisfy a number of consistency
conditions. First of all, $R$ must be an $R-$matrix, i~e.\
\begin{equation}
\Delta' (a) = R \; \Delta(a) R^{-1},
\end{equation}
where $\Delta'=\Delta$ is the twisted comultiplication.  Secondly, the
counit must satisfy
\begin{equation}
(\varepsilon \otimes id) \circ \Delta = id = (id \otimes \varepsilon)
\circ \Delta
\end{equation}
and $\Phi$ must obey
\begin{equation}
(id \otimes id \otimes \Delta) (\Phi) \; (\Delta \otimes id \otimes
id)(\Phi) = (\bbbone \otimes \Phi)\; (id \otimes \Delta \otimes
id)(\Phi)\; (\Phi \otimes \bbbone)
\end{equation}
\begin{equation}
m \Bigl((id \otimes \varepsilon \otimes id)(\Phi)\Bigr) = \bbbone .
\end{equation}
Finally, in order to be a Hopf-algebra, the antipode must have the
following properties: there exist algebra elements $\alpha$ and
$\beta$, such that
\begin{equation}
\label{Scons1}
\sum S(\Delta^{(1)}(a))\; \alpha\; \Delta^{(2)}(a) =
\varepsilon(a) \; \alpha \hspace{1cm}
\sum \Delta^{(1)}(a) \; \beta \; S(\Delta^{(2)}(a)) =
\varepsilon(a) \; \beta,
\end{equation}
where
\begin{equation}
\Delta(a) = \sum \Delta^{(1)} (a) \otimes \Delta^{(2)} (a).
\end{equation}
Furthermore, writing
\begin{equation}
\Phi=\sum \Phi_{(1)} \otimes \Phi_{(2)} \otimes \Phi_{(3)}
\hspace{1cm}
\Phi^{-1}= \sum \Phi^{-1}_{(1)} \otimes \Phi^{-1}_{(2)}
\otimes \Phi^{-1}_{(3)}
\end{equation}
$S,\; \alpha$ and $\beta$ must satisfy
\begin{equation}
\label{Scons2}
\sum \Phi_{(1)} \; \beta \; S(\Phi_{(2)})\; \alpha \; \Phi_{(3)} = \bbbone
\hspace{1cm}
\sum S(\Phi^{-1}_{(1)})\; \alpha\; \Phi^{-1}_{(2)}\; \beta\;
S(\Phi^{-1}_{(3)}) = \bbbone.
\end{equation}

I have given all these consistency conditions for completeness.  In
the following, I shall not worry about the antipode-properties, as
these do not seem to play an important r\^{o}le in the present
application. This is also justified by a theorem of Drinfel'd
\cite[Proposition 1.4]{D1}, which --- applied to the present situation ---
asserts, that there exist $\alpha$ and $\beta$ in $A_{g,t}$, such that
$S$ defined by (\ref{antip}) satisfies (\ref{Scons1},\, \ref{Scons2}).
\medskip

As far as I am aware of the literature, there is no explicit formula
for $\Phi$. However, as in \cite{D1}, one can define $\Phi$ implicitly
as follows.  Let $\psi_{i},\; i=0,1,2,3$ be four highest weight
vectors in irreducible positive energy representations $\pi_{i},\;
i=0,1,2,3$ of $\hat{g}$. I want to consider the ``chiral four point
function''
\begin{eqnarray}
{\displaystyle W_{\psi_{0}}(\psi_{1}\otimes\psi_{2}\otimes\psi_{3};
z_{1}, z_{2}, z_{3}) } & = & {\displaystyle
\lim_{z_{0}\rightarrow\infty} z_{0}^{2 \Delta_{0}} \left\langle
V(\psi_{0},z_{0}) V(\psi_{1},z_{1}) V(\psi_{2},z_{2})
V(\psi_{3},z_{3}) \right\rangle}
\nonumber \\
& = & {\displaystyle \left\langle \psi_{0}, V(\psi_{1},z_{1})
V(\psi_{2},z_{2}) V(\psi_{3},z_{3}) \right\rangle}.
\end{eqnarray}
(In the following I shall sometimes suppress the $\psi_{i}-$ or the
$z_{i}-$dependence of $W_{\psi_{0}}$ in order to stress which
dependence I have primarily in mind.)  The function
$W_{\psi_{0}}(z_{1}, z_{2}, z_{3})$ satisfies
the Knizhnik-Zamolodchikov equation \cite{KZ}, namely
\begin{equation}
\label{KZ}
\frac{\partial}{\partial z_{i}}
W_{\psi_{0}}(\psi_{1}\otimes\psi_{2}\otimes\psi_{3};
z_{1},z_{2},z_{3}) = - \bar{h}
\sum_{j\neq i}
\frac{t_{ij}}{z_{i}-z_{j}}
W_{\psi_{0}}(\psi_{1}\otimes\psi_{2}\otimes\psi_{3};z_{1},z_{2},z_{3}) ,
\end{equation}
where $\bar{h}=\frac{h}{2 \pi i}= \frac{1}{k+h^{*}}$ and
$t_{12}=t\otimes\bbbone$, $t_{23}=\bbbone\otimes t$ and similarly for
$t_{13}$. (Here, $i\in\{1,2,3\}$ and the summation in (\ref{KZ})
extends over $j=1,2,3$. This
equation follows form the KZ-equation for the four point function by
letting $z_{0}\rightarrow\infty$ as above.) These equations are
consistent, as the curvature of the corresponding connection is zero
\cite{KZ,D1}.
\smallskip

\noindent I can write the solution as
\begin{equation}
\label{ansatz}
\left( z_{1} - z_{3} \right) ^{- \bar{h} ( t_{12} + t_{13} + t_{23})}
G(\psi_{1}\otimes\psi_{2}\otimes\psi_{3};x),
\end{equation}
where $x$ is the anharmonic ratio
\begin{equation}
\label{anra}
x=\frac{z_{1}-z_{2}}{z_{1}-z_{3}} .
\end{equation}
Then $G$ has to satisfy the equation
\begin{equation}
\label{ode}
\frac{d}{dx} G(\psi_{1}\otimes\psi_{2}\otimes\psi_{3};x) =
- \bar{h} \left( \frac{t_{12}}{x} + \frac{t_{23}}{x-1} \right)
G(\psi_{1}\otimes\psi_{2}\otimes\psi_{3};x) .
\end{equation}

As shall become clear from the analysis of section~4 the different
ways of bracketing a triple tensor product correspond to the
different limits in which two of the three points involved are made to
coincide. Thus, in particular, the bracketing $\left( (\psi_{1}\otimes
\psi_{2} ) \otimes \psi_{3} \right)$ corresponds to the limit in which
$|z_{1}-z_{2}|<<|z_{2}- z_{3}|$, i.~e.\ to the limit
$x \rightarrow 0$. In this limit, $W_{\psi_{0}}$ has the expansion
\begin{equation}
\label{exp1}
W_{\psi_{0}}(\psi_{1}\otimes\psi_{2}\otimes\psi_{3}; z_{i}) \sim
\left(z_{1}-z_{2}\right)^{-\bar{h} t_{12}}
\left(z_{1}-z_{3}\right)^{-\bar{h}(t_{13}+t_{23})}
W_{\psi_{0}}^{1}(\psi_{1}\otimes\psi_{2}\otimes\psi_{3})
\end{equation}
and thus $G$ has the expansion
\begin{equation}
\label{exp1'}
G(\psi_{1}\otimes\psi_{2}\otimes\psi_{3};x) \sim x^{-\bar{h} t_{12}}
W_{\psi_{0}}^{1}(\psi_{1}\otimes\psi_{2}\otimes\psi_{3}).
\end{equation}
Expanding the right-hand-side in terms of the irreducible
subrepresentations which are contained in the tensor product
$\psi_{1}\otimes\psi_{2}$, it can be rewritten as
\begin{equation}
\sum_{p} x^{-\bar{h}\; (C_{p} - C_{1} - C_{2})} \;
W_{\psi_{0},p}^{1}(\psi_{1}\otimes\psi_{2}\otimes\psi_{3}),
\end{equation}
where $C_{i}$ is the value of the quadratic Casimir in the
representation $\pi_{i}$.

Similarly, the bracketing
$\left(\psi_{1}\otimes ( \psi_{2} \otimes \psi_{3}) \right)$
corresponds to  $|z_{2}-z_{3}|<<|z_{1}- z_{3}|$ and thus to
$x\rightarrow 1$. In this limit we have
\begin{equation}
\label{exp2}
W_{\psi_{0}}(\psi_{1}\otimes\psi_{2}\otimes\psi_{3}; z_{i})
\sim \left(z_{2}-z_{3}\right)^{-\bar{h} t_{23}}
\left(z_{1}-z_{3}\right)^{-\bar{h}(t_{12}+t_{13})}
W_{\psi_{0}}^{2}(\psi_{1}\otimes\psi_{2}\otimes\psi_{3}).
\end{equation}
and thus
\begin{equation}
\label{exp2'}
G(\psi_{1}\otimes\psi_{2}\otimes\psi_{3};x)
\sim (1-x)^{-\bar{h} t_{23}}
W_{\psi_{0}}^{2}(\psi_{1}\otimes\psi_{2}\otimes\psi_{3}).
\end{equation}
Again, the right-hand-side can be rewritten as
\begin{equation}
\sum_{r} (1-x)^{-\bar{h}\; (C_{r} - C_{2} - C_{3})}\;
W_{\psi_{0},r}^{2}(\psi_{1}\otimes\psi_{2}\otimes\psi_{3}),
\end{equation}
where $r$ parametrises the irreducible subrepresentations in the
tensor product $\psi_{2}\otimes\psi_{3}$.

Obviously, there is some freedom in the normalisation of the four
point functions. However, once we have chosen a normalisation for say
$W_{\psi_{0}}^{1}$, $W_{\psi_{0}}^{2}$ is uniquely determined by
analytically continuing the solution corresponding to
$W_{\psi_{0}}^{1}$. In particular, because of (\ref{KZ}) and
(\ref{ode}), we can find an invertible element
\begin{equation}
\Phi\in U(g)\otimes U(g)\otimes U(g)
\end{equation}
such that
\begin{equation}
W_{\psi_{0}}^{1}(\psi_{1}\otimes\psi_{2}\otimes\psi_{3}) =
W_{\psi_{0}}^{2}\Bigl(\Phi\left(\psi_{1}\otimes\psi_{2}\otimes\psi_{3}
\right)\Bigr)
\end{equation}
independent of $\psi_{0}$ and $z_{i}$. (Formally, one can write $\Phi$
as an ordered exponential in the Lie algebra generators, integrating
up (\ref{ode}).) $\Phi$ is invertible and it can be shown --- using
conformal field theory arguments --- that it satisfies all of the above
consistency conditions \cite{D1}.

I would like to point out that for generic $q$ the matrix elements of
$\Phi$ can be expressed in terms of the coupling constants of the
theory. I shall use this fact implicitly in section~5.
\smallskip

\section{The r\^{o}le as internal symmetry}
\renewcommand{\theequation}{4.\arabic{equation}}
\setcounter{equation}{0}

In section~2, I have shown that the (holomorphic) chiral theory
corresponding to a WZW-model is given as
\begin{equation}
{\cal H}_{chir} =\bigoplus_{l} {\cal H}_{l} \otimes \overline{W}_{l}.
\end{equation}
The chiral algebra ${\cal A}=\hat{g}$ acts on ${\cal H}_{chir}$ as
\begin{equation}
\label{chirala}
a \mapsto a\otimes\bbbone
\end{equation}
and the vacuum representation is the summand in ${\cal H}_{chir}$
corresponding to
\begin{equation}
{\cal A}\; \Omega \otimes \overline{\Omega}.
\end{equation}
I would like to remark, that the whole vertex operators corresponding
to this representation contain the holomorphic currents
\begin{equation}
J^{a}(z) = \phi(J^{a}_{-1}\Omega\otimes\overline{\Omega}; z,\bar{z}) ,
\end{equation}
as the right-hand-side is in fact independent of $\bar{z}$. Thus the
holomorphic currents do belong to the (holomorphic) chiral theory.
\medskip

There is a natural action of the quasi quantum group $A_{g,t}$ on this
Hilbert space, given by
\begin{equation}
\label{symma}
\bar{a} \mapsto \bbbone\otimes \bar{a},
\end{equation}
where $\bar{a}$ is regarded as an element of the horizontal Lie
algebra $g$ of the affine algebra $\overline{\cal A}=\hat{g}$.  As
mentioned before, the comultiplication formula \cite{MG1} restricted
to this subalgebra is trivial and thus agrees with (\ref{com}).
Furthermore, the braiding of the chiral vertex operators is described
by the $R$-matrix of the quasi quantum group and thus the universal
enveloping algebra of the horizontal Lie algebra is $A_{g,t}$ rather
than $U(g)$.  To explain this in more detail, recall that the whole
vertex operators commute, i.~e.\
\begin{equation}
\label{wholcom}
\phi(\psi\otimes\overline{\psi};z,\bar{z})\;
\phi(\chi\otimes\overline{\chi};\zeta,\bar{\zeta}) =
\phi(\chi\otimes\overline{\chi};\zeta,\bar{\zeta})\;
\phi(\psi\otimes\overline{\psi};z,\bar{z})
\end{equation}
upon analytic continuation of $z$ and $\zeta$, and correspondingly
``anti-analytic continuation'' of $\bar{z}$ and $\bar{\zeta}$.
\footnote{Because of (\ref{decoup}) analytic and
anti-analytic continuation is a well-defined concept for correlation
functions. The above assertion means that this is true for all
correlation functions.} Since I have explicitly constructed chiral
vertex operators as a certain limit of whole vertex operators, I can
calculate the braiding as follows:
\begin{Theorem}
Upon anticlockwise analytic continuation of the left hand side one has
\begin{equation}
\label{Theorem1}
V(\psi\otimes\bar{v},z) \; V(\chi\otimes \bar{w},\zeta)\;
\left(\Omega \otimes \overline{\Omega}\right)=
\sum_{\bar{w}', \bar{v}'} {R \;}_{\bar{v}'\;\bar{w}'}^{\bar{v}\;\bar{w}}
\; \; V(\chi\otimes\bar{w}',\zeta) \; V(\psi\otimes \bar{v}',z)
\; \left(\Omega \otimes \overline{\Omega}\right),
\end{equation}
where ${R \;}_{\bar{w}'\;\bar{v}'}^{\bar{w}\;\bar{v}}$ is the matrix
element of the $R-$matrix (\ref{rm}) with $q=e^{\frac{\pi
i}{k+h^{*}}}$ and the sum extends over a basis of the corresponding
finite dimensional internal parameter spaces.
\end{Theorem}
\medskip

\noindent {\em Proof:} To calculate the braiding of the
analytic continuation, consider
the scalar product with $(\varphi\otimes\bar{u})\in {\cal H}_{l}\otimes
\overline{W}_{l}$. The correlation function of the corresponding
whole vertex operators satisfies
\begin{equation}
\label{decopf}
\left\langle (\varphi\otimes\bar{u}),
\phi(\psi\otimes\bar{v}; z,\bar{z})\;
\phi(\chi\otimes\bar{w}; \zeta,\bar{\zeta})\;
\left(\Omega \otimes \overline{\Omega}\right) \right\rangle =
(\bar{z} - \bar{\zeta}) ^{\bar{\Delta}_{l} - \bar{\Delta}_{\bar{v}} -
\bar{\Delta}_{\bar{w}}} \; F_{\psi \chi}(z,\zeta).
\end{equation}
Upon anticlockwise analytic continuation of $z$ around $\zeta$ (which
corresponds to clockwise analytic continuation of $\bar{z}$ around
$\bar{\zeta}$)  we obtain
\begin{equation}
e^{-\pi i (\bar{\Delta}_{l} - \bar{\Delta}_{\bar{v}} -
\bar{\Delta}_{\bar{w}})}\; (\bar{\zeta} - \bar{z})^{\bar{\Delta}_{l} -
\bar{\Delta}_{\bar{v}} - \bar{\Delta}_{\bar{w}}}
\widetilde{F}_{\psi \chi}(z,\zeta)
\end{equation}
which must equal
\begin{equation}
(\bar{\zeta} - \bar{z}) ^{\bar{\Delta}_{l} - \bar{\Delta}_{\bar{v}} -
\bar{\Delta}_{\bar{w}}} F_{\chi \psi}(\zeta,z),
\end{equation}
as the whole vertex operators satisfy (\ref{wholcom}).

We obtain the correlation function of the corresponding chiral vertex
operators by multiplying the above expression with
$(\bar{\zeta}-\bar{z})^{-(\bar{\Delta}_{l} - \bar{\Delta}_{\bar{v}}-
\bar{\Delta}_{\bar{w}})}$ --- we do not have to take the limit
$\bar{\zeta}\rightarrow \bar{z}$, as we have already projected onto the
lowest $\bar{L}_{0}-$eigenspace by taking the scalar product.  Thus
the summand in the operator product expansion (of the two chiral
vertex operators) corresponding to ${\cal H}_{l}\otimes
\overline{W}_{l}$ exhibits the braiding phase
\begin{equation}
\label{braid}
e^{\pi i (\bar{\Delta}_{l} - \bar{\Delta}_{\bar{v}} -
\bar{\Delta}_{\bar{w}})}.
\end{equation}
However, because of \cite{MG1}, the fusion of $\overline{W}_{\bar{v}}$
and $\overline{W}_{\bar{w}}$ is just the ordinary tensor product,
since the comultiplication is trivial. Thus, to pick out the summand
${\cal H}_{l}\otimes \overline{W}_{l}$, we only have to decompose the
tensor product of $\overline{W}_{\bar{v}}$ and
$\overline{W}_{\bar{w}}$ into the direct sum of irreducible
representations and project onto the summand $\overline{W}_{l}$.
However, by construction of the $R-$matrix, the matrix element of $R$
in this subspace is exactly (\ref{braid}) and thus the theorem is
proved.
\medskip

I would like to remark, that the tensor product of
$\overline{W}_{\bar{v}}$ and $\overline{W}_{\bar{w}}$ contains in
general irreducible representations, which do not appear in the
tensor product of the corresponding current algebra because of
truncation. If $\overline{W}_{l}$ is such a
representation, $F_{\psi,\chi}$ in (\ref{decopf}) is identically zero,
as the $\phi$'s are vertex operators of a well-defined conformal
field theory.  Hence the operator product expansion of the two chiral
vertex operators does not contain the corresponding conformal family
and in particular the three-point-function of both sides of
(\ref{Theorem1}) vanishes identically. This also implies that the
$R$-matrix given above is in general not uniquely determined.
\medskip

The above theorem describes the braiding in a rather special case,
namely when the product of vertex operators acts on the vacuum.
The general case can be derived from this special case, once
the duality properties of the chiral theory have been established.
In general, as we shall see, the chiral theory does not satisfy
the duality property of the whole theory
\begin{equation}
\label{duality}
\phi(\psi\otimes\bar{v}; z,\bar{z}) \;
\phi(\chi\otimes\bar{w}; \zeta,\bar{\zeta})\;  = \;
\phi\left( \phi\left(\psi\otimes\bar{v};
z-\zeta,\bar{z}-\bar{\zeta}\right)
(\chi\otimes\bar{w}); \zeta,\bar{\zeta}\right),
\end{equation}
but only a modified version. (This is due to the fact that the
different limit procedures (in the definition of the chiral vertex
operators) do not commute.)
In particular, this implies, that the
braiding is not really local. Indeed, the braiding
of two chiral vertex operators will turn out to depend also on the
state the product of vertex operators is acting on (\ref{rgen}).

The fact that the duality property has to be
modified is intimately related to the fact that the $R$-matrix
given above does not satisfy the quantum group consistency condition
(\ref{rmc}), but only the weaker conditions (\ref{rc1}, \ref{rc2}).
To complete my argument that the universal enveloping
algebra of the horizontal Lie algebra is really $A_{g,t}$, I have to
show that the chiral vertex operators satisfy a weaker duality
property related to $\Phi$.

\begin{Theorem}
\begin{equation}
\label{th2}
V \Bigl( V(\psi\otimes\bar{v},z-\zeta) \; (\chi\otimes\bar{w}),
\zeta \Bigr) \; (\varphi \otimes \bar{u}) =
\sum_{\bar{v}', \bar{w}', \bar{u}'}
{\Phi \;}_{\bar{v}'\;\bar{w}'\;\bar{u}'}^{\bar{v}\; \bar{w}\; \bar{u}
}\;\; V(\psi\otimes\bar{v}',z) \; V(\chi\otimes\bar{w}',\zeta)\;
(\varphi\otimes \bar{u}') ,
\end{equation}
where ${\Phi \;}_{\bar{v}'\; \bar{w}'\; \bar{u}'}^ {\bar{v}\;
\bar{w}\; \bar{u}}$ are the matrix elements of $\Phi$ and the sum
extends over a basis of the finite dimensional internal parameter
spaces.
\end{Theorem}
\medskip

\noindent {\em Proof:} As above it is sufficient to consider the
case, where we take the
scalar product with $(\omega\otimes \bar{x}) \in {\cal H}_{l} \otimes
\overline{W}_{l}$. For the corresponding whole vertex operators we have
$$
\left\langle (\omega\otimes \bar{x}),
\phi\left( \phi(\psi\otimes\bar{v}; z-\zeta,\bar{z}-\bar{\zeta})
\; (\chi\otimes\bar{w}); \zeta,\bar{\zeta}\right)
(\varphi\otimes\bar{u}) \right\rangle
\hspace*{4.5cm}
$$
\begin{equation}
\label{vier1}
\hspace*{4cm} = \sum_{p}
T_{p}(\omega, \psi, \chi, \varphi; z, \zeta) \;
\widehat{W}_{\bar{x},\bar{p}}^{1}
(\bar{v} \otimes \bar{w} \otimes \bar{u}; \bar{z},
\bar{\zeta}, 0),
\end{equation}
and
$$
\left\langle (\omega\otimes \bar{x}),
\phi(\psi\otimes\bar{v}; z,\bar{z}) \;
\phi(\chi\otimes\bar{w}; \zeta,\bar{\zeta}) \;
(\varphi\otimes\bar{u}) \right\rangle
\hspace*{4.5cm}
$$
\begin{equation}
\label{vier2}
\hspace*{4cm} = \sum_{r}
S_{r}(\omega, \psi, \chi, \varphi; z, \zeta) \;
\widehat{W}_{\bar{x},\bar{r}}^{2}
(\bar{v} \otimes \bar{w} \otimes \bar{u}; \bar{z},
\bar{\zeta}, 0),
\end{equation}
where the indices $p, \bar{p}$ and $r,\bar{r}$ indicate the different
conformal families, which contribute in the $t$- and $s$-channel,
respectively. (\ref{vier1}) is defined for $|\zeta| > |z-\zeta|$,
(\ref{vier2}) for $|z| > |\zeta|$ and the right-hand sides of
(\ref{vier1}) and (\ref{vier2}) are the same analytic function in the
four (independent) variables $z,\zeta,\bar{z}$ and $\bar{\zeta}$. The
``equality'' of (\ref{vier1}) and (\ref{vier2}) is the duality property
of the whole theory (\ref{duality}), which
can be established from first principles using locality
\cite{Peter89}.

To obtain the left-hand-side of
(\ref{th2}) we have to take the limit $(\bar{z}-\bar{\zeta})\rightarrow 0$
and $\bar{\zeta} \rightarrow 0$ in
(\ref{vier1}). Using the above equality this is the same as
\begin{equation}
\sum_{p} T_{p}(\omega, \psi, \chi, \varphi; z, \zeta) \;
\lim_{\bar{\zeta} \rightarrow 0} \;\bar{\zeta}^{
\Delta_{\bar{u}} + \Delta_{\bar{p}} - \Delta_{\bar{x}}} \;
\lim_{\bar{z}\rightarrow \bar{\zeta}}\;
(\bar{z} - \bar{\zeta})^{\Delta_{\bar{v}} +
\Delta_{\bar{w}} - \Delta_{\bar{p}}} \;
\widehat{W}_{\bar{x},\bar{p}}^{1} (\bar{z}, \bar{\zeta},0).
\end{equation}
In the notation of section~3 we obtain thus
\begin{equation}
\label{expre1}
\sum_{p} T_{p}(\omega, \psi, \chi, \varphi; z, \zeta) \;
W_{\bar{x},\bar{p}}^{1}(\bar{v}\otimes\bar{w}\otimes\bar{u}).
\end{equation}

The right-hand-side of (\ref{th2}) is obtained by
taking the limit $\bar{\zeta}\rightarrow 0$ and
$\bar{z}\rightarrow 0$ in (\ref{vier2}).
Using again the above equality, this is the same as
\begin{equation}
\sum_{r}
S_{r}(\omega, \psi, \chi, \varphi; z, \zeta)
\lim_{\bar{z} \rightarrow 0} \;\bar{z}^{
\Delta_{\bar{v}} + \Delta_{\bar{r}} - \Delta_{\bar{x}}} \;
\lim_{\bar{\zeta}\rightarrow 0} \;
\bar{\zeta}^{\Delta_{\bar{w}} +
\Delta_{\bar{u}} - \Delta_{\bar{r}}} \;
\widehat{W}_{\bar{x},\bar{r}}^{2}(\bar{z},\bar{\zeta},0),
\end{equation}
and thus in the notation of section~3 equals
\begin{equation}
\label{expre2}
\sum_{r}
S_{r}(\omega, \psi, \chi, \varphi; z, \zeta)
W_{\bar{x},\bar{r}}^{2}(\bar{v}\otimes\bar{w}\otimes\bar{u}).
\end{equation}
For fixed $z$ and $\zeta$, the function (\ref{vier1}) and
(\ref{vier2}) satisfies the KZ-equation for $\bar{z}$ and
$\bar{\zeta}$. Therefore, the two expressions (\ref{expre1}) and
(\ref{expre2}) are related as in section~3, and thus, by
definition of $\Phi$, the theorem holds.
\medskip

Putting the information of Theorem~1 and Theorem~2 together, we can
now determine the braiding of two chiral vertex operators in the
general case. One easily finds that the anticlockwise analytic
continuation of $z$ around $\zeta$ on the left-hand-side
equals
\begin{equation}
\label{rgen}
V(\psi\otimes\bar{v},z) \; V(\chi\otimes \bar{w},\zeta)\;
(\varphi\otimes \bar{u}) =
\sum_{\bar{w}', \bar{v}', \bar{u}'}
\left(\Phi_{213} \; R_{12} \Phi^{-1} \right)_{\bar{v}'\; \bar{w}'\; \bar{u}'}
^ {\bar{v}\;\bar{w}\; \bar{u}}
\; \; V(\chi\otimes\bar{w}',z) \; V(\psi\otimes \bar{v}',\zeta)
\;(\varphi\otimes \bar{u}'),
\end{equation}
where $(\Phi_{213} \; R_{12} \Phi^{-1})_{\bar{v}'\; \bar{w}'\; \bar{u}'}
^ {\bar{v}\;\bar{w}\; \bar{u}}$ is the matrix element of the
composition of the three maps. Thus in particular the braiding
of two chiral vertex operators also depends on the state the
operator product is acting on.
\bigskip

I have thus shown that the universal enveloping algebra of the
anti-holomorphic horizontal Lie algebra forms indeed $A_{g,t}$. Its
physical significance is that it plays the r\^{o}le of an internal
symmetry for the chiral WZW-theory, as it satisfies the following
properties:
\begin{itemize}
\item[] it commutes with the chiral algebra ${\cal A}=\hat{g}$,
which can be interpreted to be some sort of observable algebra,
\item[] it annihilates the vacuum $\Omega\otimes\overline{\Omega}$,
as $\bar{a} \; \overline{\Omega} =0$ for all $\bar{a}\in g\subset
\overline{\cal A}$
\item[] as a consequence of the comultiplication formula for the
horizontal subalgebra \cite{MG1} chiral vertex operators transform
covariantly under $A_{g,t}$, i.~e.\
\begin{equation}
\label{cova}
\bar{a} \; V(\psi\otimes\bar{w},z) = V(\psi\otimes\bar{w},z)\; \bar{a}
+ V(\psi\otimes \bar{a} \bar{w},z)
\hspace*{1.5cm} \mbox{for all $\bar{a}\in g\subset \overline{\cal A}$}
\end{equation}
\item[]  the braiding (Theorem 1) and the duality (Theorem 2) of
the chiral vertex operators are described by the $R$-matrix and $\Phi$
of $A_{g,t}$, respectively.
\end{itemize}
\bigskip

The quasi Hopf algebra is not coassociative and therefore the
chiral vertex operators only satisfy a modified duality property.
For generic $q$, the duality property of the chiral vertex operators
can be restored, using the isomorphism between $A_{g,t}$ and the
coassociative quantum group $U_{q}(g)$. However, for the physically
relevant case of integer $k$ this construction has to be modified.
In particular, the duality property of the chiral theory can not
be completely restored, as the quantum group has to be replaced by the
corresponding truncated quasi quantum group. This seems to indicate
that a chiral theory possesses in general only a weaker version of duality.

\section{The quantum group $U_{q}(g)$}
\renewcommand{\theequation}{5.\arabic{equation}}
\setcounter{equation}{0}

It has been shown by Drinfel'd \cite{D2}, that the quasi quantum group
defined above and the (well-known) quantum group $U_{q}(g)$ are
isomorphic for generic $q$, where $h$ and $q$ are related as in
section~2, namely
\begin{equation}
\label{qrel}
q=e^{h/2}= e^{\frac{\pi i}{k + h^{*}}}.
\end{equation}
Here I choose the conventions as in \cite{Majid}, i.~e.\
\begin{equation}
[ H_{i}, X_{\pm j} ] = \pm (\alpha_{i}, \alpha_{j}) X_{\pm j},
\hspace{2cm}
[X_{+i}, X_{- j}] = \delta_{ij} \frac{ q^{H_{i}} - q^{- H_{i}}} {q -
q^{-1}}
\end{equation}
and
\begin{equation}
\Delta_{q} (H_{i}) = H_{i}\otimes\bbbone + \bbbone\otimes H_{i},
\hspace{2cm}
\Delta_{q} ( X_{\pm i}) = X_{\pm i} \otimes q^{\frac{H_{i}}{2}} +
q^{- \frac{H_{i}}{2}} \otimes X_{\pm i}.
\end{equation}

By this I mean that there exists an invertible map $\varphi: U_{q}(g)
\rightarrow A_{g,t}$, $\varphi=id\; (mod\; h)$ and an invertible
element $F\in U(g)\otimes U(g)$, $F=\bbbone\otimes\bbbone \; (mod\;
h)$, such that
\begin{equation}
\label{twist}
F\; \Delta(\varphi(a)) = (\varphi\otimes \varphi)\,
\Delta_{q}(a)\; F
\end{equation}
for all $a\in U_{q}(g)$. Furthermore the $R-$matrices are related by
\begin{equation}
\label{rmt}
\left(\varphi\otimes\varphi\right) \left(R_{q}\right) = F' \; R \; F^{-1},
\end{equation}
where $F'=F_{21}$ in the usual notation, and
\begin{equation}
\label{phirel}
\Phi_{q}=
\Bigl(\bbbone\otimes\bbbone\otimes\bbbone\Bigr) =
\left(\bbbone\otimes F \right)\;  \left( id \otimes \Delta \right) (F)
\; \Phi\; \Bigl(\left( \Delta\otimes id \right) (F) \Bigr)^{-1} \;
\left( F \otimes \bbbone \right)^{-1}.
\end{equation}
In particular this implies, that the quasi quantum group with
non-trivial $\Phi$ is in fact isomorphic to a quantum group with
$\Phi_{q}=\bbbone\otimes\bbbone\otimes\bbbone$.
\smallskip

To be more specific I want to construct this isomorphism explicitly
for the case of $su(2)$. I define the map (analogously to \cite{CZ})
\begin{equation}
\varphi: U_{q}(su(2)) \rightarrow A_{su(2),t}
\end{equation}
by
\begin{equation}
\varphi(H)=H \hspace{2.0cm}
\varphi(X_{\pm}) = X_{\pm} \; P_{\pm},
\end{equation}
where
\begin{equation}
P_{\pm}=\sqrt{ \frac{ [ j \mp m]_{q}\; [j \pm m + 1 ]_{q}} {(j\mp m)
(j \pm m + 1) } },
\end{equation}
and
\begin{equation}
\left[ p \right] _{q} : = \frac{ q^{p} - q^{-p}}{q-q^{-1}}.
\end{equation}
Here, $j (j+1)$ is the eigenvalue of the quadratic Casimir
\begin{equation}
C:= \frac{1}{4} H^{2} + \half \left( X_{+}\, X_{-} + X_{-}\, X_{+}
\right)
\end{equation}
and $m$ is the eigenvalue of $\half H$. Thus I can express $j$ and
$m$ in terms of elements of $A_{su(2),t}$. This map is well-defined
for arbitrary $q$. For generic $q$ it is also invertible.
\medskip

Next I want to relate the action of the two algebras on tensor
products.  To this end I define the map $\widetilde{F}$
\begin{equation}
\widetilde{F}: V_{j_{1}} \otimes V_{j_{2}} \rightarrow
V_{j_{1}} \otimes_{q} V_{j_{2}}
\end{equation}
by
\begin{eqnarray}
{\displaystyle
\widetilde{F} \Bigl( |j_{1}, m_{1} \rangle \otimes | j_{2}, m_{2}
\rangle \Bigr)}
& := & {\displaystyle
\sum_{m_{1}', m_{2}'} \sum_{J} \;\;\left(d^{J}_{j_{1},j_{2}}\right)^{-1}\;\;
\langle j_{1} j_{2} m_{1}' m_{2}' | J, m_{1}+m_{2} \rangle_{q}}
\hspace*{3cm}
\nonumber \\
& & \hspace*{0.8cm}
\label{fdef}
{\displaystyle
\langle j_{1} j_{2} m_{1} m_{2} | J, m_{1}+m_{2} \rangle \;\;\;
|j_{1}, m_{1}'\rangle \otimes_{q} | j_{2}, m_{2}'\rangle,}
\end{eqnarray}
where $d^{J}_{j_{1},j_{2}}$ are constants to be specified shortly,
$\langle j_{1} j_{2} m_{1} m_{2} | J,M \rangle$ and $\langle j_{1}
j_{2} m_{1} m_{2} | J,M \rangle_{q}$ are the Clebsch-Gordon
coefficients of $su(2)$ and $U_{q}(su(2))$, respectively, and the
subscript $q$ of the tensor product indicates, that it is to be
regarded as a representation of the quantum group via the action
$(\varphi\otimes\varphi) \Delta_{q}(a)$.  Then, by construction,
$\widetilde{F}$ satisfies
\begin{equation}
\widetilde{F}\; \Delta(\varphi(a)) = (\varphi\otimes \varphi)\,
\Delta_{q}(a)\; \widetilde{F}.
\end{equation}
Thus, I can use the Racah formula \cite{Messiah} for the
Clebsch-Gordon coefficients of $su(2)$
and a similar formula for $U_{q}(su(2))$ \cite{KR}
to rewrite (for generic $q$) $\widetilde{F}$
as $F\in U(g)\otimes U(g)$.  $F$ satisfies
then by construction (\ref{twist}) and (\ref{rmt}). Furthermore,
$F$ is the identity modulo $h$, if
$d^{J}_{j_{1},j_{2}} \rightarrow 1$ as $k\rightarrow \infty$.
\smallskip

A priori it is not clear, whether $F$ satisfies (\ref{phirel}).
However, by Schur's lemma, $F$ can differ from the invertible element
relating $A_{su(2),t}$ and $U_{q}(su(2))$ (for generic $q$) at most by
a scalar function of $q$ for each irreducible subrepresentation in the
tensor product of two (irreducible) representations. I can therefore
choose the ``coupling constants'' $d^{J}_{j_{1},j_{2}}$ so
that $F$ satisfies (\ref{phirel}).

To fix these constants explicitly, I shall use results obtained from the
Coulomb gas representation of the $su(2)$ conformal field theory
\cite{DF1,CF,FZ1,Do1,Do2}. In particular, the coupling constants
for the whole theory turn out to be \cite{Do2}
\begin{equation}
\label{wholecoup}
D^{j_{3},m_{3},\overline{m}_{3}}_{j_{1},m_{1},
\overline{m}_{1};j_{2},m_{2}, \overline{m}_{2}} =
\sqrt{\frac{(2 j_{1} + 1) (2 j_{2} + 1)}{(2 j_{3} + 1)}} \;
\langle j_{1} j_{2} m_{1} m_{2} | j_{3},m_{3} \rangle \;
\langle j_{1} j_{2} \overline{m}_{1} \overline{m}_{2} | j_{3},
\overline{m}_{3} \rangle
\; d^{j_{3}}_{j_{1},j_{2}},
\end{equation}
where
\begin{equation}
\label{ddef}
d^{j_{3}}_{j_{1},j_{2}} = \frac{a_{j_{1}} a_{j_{2}}}{a_{j_{3}}}\;
\prod_{l=1}^{j_{1} + j_{2} - j_{3}}
\frac{\Gamma(\frac{l}{k+2})} {\Gamma(-\frac{l}{k+2})} \;
\prod_{l=0}^{j_{1} + j_{2} - j_{3}-1}
\frac{\Gamma(-\frac{ 2 j_{1} - l}{k+2})\; \Gamma(-\frac{ 2 j_{2} - l}{k+2})
\; \Gamma(\frac{ 2 j_{3} +2 + l}{k+2})}{\Gamma(\frac{ 2 j_{1} - l}{k+2})\;
\Gamma(\frac{ 2 j_{2} - l}{k+2})\; \Gamma(-\frac{ 2 j_{3} +2 +
l}{k+2})}
\end{equation}
and
\begin{equation}
a_{j}= \left[ \prod_{l=1}^{2 j} \frac{\Gamma(\frac{l}{k+2})\;
\Gamma(-\frac{1+l}{k+2})}{\Gamma(-\frac{l}{k+2})\;
\Gamma(\frac{1+l}{k+2})}\right]^{\half}.
\end{equation}
The square root in (\ref{wholecoup}) is a consequence of the
normalisation convention of \cite{DF2}. $d^{j_{3}}_{j_{1},j_{2}}$
is the $k$-dependent part and tends to $1$ as $k\rightarrow \infty$.

Using the explicit expression for the coupling constants of the
whole theory, we can calculate the analytic continuation of the
left-hand-side of (\ref{th2}) and thus determine the
matrix elements of $\Phi$. This fixes the constants
$d^{J}_{j_{1},j_{2}}$ in (\ref{fdef}) to be equal to (\ref{ddef}).

I would like to point out that the results of the Coulomb gas
representation have only been derived for generic (irrational) $k$.
On the other hand, the matrix elements of the $R$-matrix and
the $\Phi$-map of the quasi quantum group $A_{g,t}$
depend continuously on $k$. We shall use this argument below to show
that the formula for the constant $d^{J}_{j_{1},j_{2}}$
(for suitably restricted $j_{1}, j_{2}$ and $J$) extends to
the case of a root of unity.
\medskip

The above formula for $\varphi$ is only invertible and
$\widetilde{F}$ is only well defined on all tensor products
if $q$ is generic. The breakdown of the
formulae at a root of unity is mirrored by the
fact that the quantum group ceases to be semisimple at roots of
unity, as not all representations of the quantum group are
completely reducible. Thus, at a root of unity, the
symmetry algebra will not be the original quantum group, but only some
modification.

In fact, at a root of unity, the quantum group has to be replaced
by its truncated version. By this I mean that one
restricts the action of the quantum
group to the so-called physical representations and considers only the
projection of the tensor product onto its completely decomposable
part. This truncated version of the quantum group has been studied in
\cite{MaS1}, where it was called the canonically associated
``weak quasi-triangular quasi Hopf algebra''. It is important to note,
that the truncation breaks the coassociativity and thus that the
resulting algebra is only a {\em quasi} Hopf algebra. However, the
corresponding $\Phi$-map can be easily determined (a formula
is given in \cite{MaS1}) and it is trivial on all
triple tensor products, which do not exhibit any truncation
\footnote{It should be noted that the coassociator $\Phi$ of a weak
quasi quantum group is not invertible, but rather possesses only
a quasi-inverse. The same applies to the $R$ matrix.}.

The physical representations are those, which correspond to
the unitary positive energy representations of the affine algebra
at level $x=k / \psi^{2}$. For $su(2)$, $x=k$ and the unitary
positive energy representations are characterised by $j\leq k/2$.
It is easy to see, that $\varphi$ is indeed invertible on these
representations.

To project the tensor product of physical representations onto its
completely decomposable part one restricts the sum
over $J$ in (\ref{fdef}) to $J\leq k - j_{1} - j_{2}$.
This truncation is necessary to give a well-defined meaning to
$\widetilde{F}$, since $[j_{1}+j_{2}+J+1]_{q}!= 0$ for
$j_{1}+j_{2}+J+1\geq k+2$ and the
Clebsch-Gordon coefficient for $U_{q}(su(2))$ becomes singular.
The truncation is precisely the
truncation of the fusion rules of the corresponding WZW-model
\cite{Walton,FGP,FD}.
\footnote{For the case of $su(n)$, this follows already from the
fact that there is only one associative fusion ring, which is a
certain truncation of the Clebsch-Gordon series of the corresponding
Lie algebra \cite{GN}.} On this truncated tensor product
the map $\widetilde{F}$ is invertible.

The modified maps do indeed relate the so truncated
quasi quantum group $A_{g,t}$
and the truncated quantum group. This follows from the fact, that
the matrix elements of $R$ and $\Phi$ and the map $\widetilde{F}$
depend continuously on $k$. Therefore on the truncated tensor product
the $R$ matrix and the $\Phi$ map of the truncated quantum group agree
with the right hand sides of (\ref{rmt}) and (\ref{phirel}).
In particular, the constants $d^{J}_{j_{1},j_{2}}$ are still given by
(\ref{ddef}).
\medskip

I can thus use these maps to regard the internal parameter spaces of the
chiral vertex operators as representation spaces of the quantum group.
By this I mean that I define new chiral vertex operators
\begin{equation}
V^{q} (\psi\otimes\bar{w},z) :{\cal H}_{chir} \rightarrow {\cal H}_{chir},
\end{equation}
whose action on a state in ${\cal H}_{chir}$ is given as
\begin{equation}
V^{q} (\psi\otimes\bar{w},z) \; (\chi\otimes\bar{v}) : =
\sum_{\bar{w}', \bar{v}'}
\left(\widetilde{F}^{-1}\right)_{\bar{w}', \bar{v}'}^{\bar{w}, \bar{v}}
V(\psi\otimes\bar{w}',z) \; (\chi\otimes\bar{v}'),
\end{equation}
where $\widetilde{F}^{-1}$ is the inverse of $\widetilde{F}$ on the
truncated tensor product and $V(\psi\otimes\bar{w}',z)$ is the
original chiral vertex operator. Similarly, the original chiral vertex
operators can be expressed in terms of the new chiral vertex operators
using $\widetilde{F}$. It should be noted that the two descriptions
are indeed in one-to-one correspondence
as the Clebsch-Gordon series of the truncated tensor
product and the fusion rules of the WZW-model agree.

All arguments of the previous section
can be easily adapted. Thus, in particular, Theorem~1 and Theorem~2
hold, where $R$ and $\Phi$ are now the $R$-matrix and the $\Phi$-map
of the truncated quantum group, and the covariance property
(\ref{cova}) becomes
\begin{equation}
a \; V^{q}(m\otimes\bar{w},z) = \sum V^{q}(m\otimes \Delta_{q}^{(1)} (a)
\bar{w},z) \; \Delta_{q}^{(2)}(a)
\end{equation}
for all $a\in U_{q}(g)$, where I have adapted the notation
\begin{equation}
\label{not}
\Delta_{q}(a) = \sum \Delta_{q}^{(1)}(a)
\otimes \Delta_{q}^{(2)}(a).
\end{equation}
This exhibits the truncated quantum group as the internal symmetry
algebra of the WZW-model for integer $k$.
\medskip

Having specified $\widetilde{F}$ explicitly, I can also calculate the
braiding matrix for all vertex operators. As an example, let us
consider the braiding matrix of two spin-$\half$ vertex operators. In
the usual notation for chiral vertex operators introduced in
\cite{MS}, the braiding matrix is given for $0 < j < \frac{k}{2}$ as
\begin{equation}
\label{braid1}
B_{p,p'}\left[
\begin{array}{cc}
\half & \half \\
j & j
\end{array}
\right] =
\left(
\begin{array}{cc}
d_{+} & 0 \\ 0 & d_{-}
\end{array} \right) \;
R_{m,m'}\left[
\begin{array}{cc}
\half & \half \\
j & j
\end{array} \right] \;
\left(
\begin{array}{cc}
d_{+}^{-1} & 0 \\ 0 & d_{-}^{-1}
\end{array} \right),
\end{equation}
and for $j=0$ or $j=\frac{k}{2}$ as
\begin{equation}
\label{braid2}
B_{p,p'}\left[
\begin{array}{cc}
 \half & \half \vspace*{0.3cm} \\
0 & 0
\end{array}
\right] \; = \; - q^{-\frac{3}{2}} \; \delta_{p,p'} \; =
B_{p,p'}\left[
\begin{array}{cc}
\half & \half \vspace*{0.3cm} \\
\frac{k}{2} & \frac{k}{2}
\end{array}
\right].
\end{equation}
Here
\begin{equation}
R_{m,m'}\left[
\begin{array}{cc}
\half & \half \\
j & j
\end{array} \right] =
\left(
\begin{array}{cc}
\frac{- q^{-(2 j + \thalf)}}{[ 2 j + 1]_{q}} &
\frac{\sqrt{q^{-1} [2 j]_{q} [ 2 j + 2]_{q}}}{[2 j + 1]_{q}}
\vspace*{0.3cm}\\
\frac{\sqrt{q^{-1} [2 j]_{q} [ 2 j + 2]_{q}}}{[2 j + 1]_{q}} &
\frac{ q^{ 2 j + \half}}{[2 j + 1]_{q}}
\end{array} \right)
\end{equation}
is the braiding matrix of $U_{q}(su(2))$ and
\begin{equation}
d_{\pm} = d^{j\pm\half}_{j,\half} .
\end{equation}
The diagonal matrices in (\ref{braid1}) relate the coupling
constants of the original theory and the chiral theory. The explicit
from follows directly from (\ref{fdef}).
(\ref{braid2}) is just a phase, as only the vacuum representation
in the tensor product of the two
spin-$\half$ representations contributes.

The expressions agree with the result given in \cite{TK}. This is
immediate for (\ref{braid2}). For the case of (\ref{braid1}) this
follows from
\begin{equation}
\frac{d_{+}}{d_{-}} =
\frac{ \Gamma(-\frac{2 j + 1}{k+2})}{\Gamma(\frac{2 j + 1}{k+2})} \;
\sqrt{\frac{\Gamma(\frac{2 j }{k+2})\; \Gamma(\frac{2 j + 2}{k+2})}
{\Gamma(-\frac{2 j }{k+2})\; \Gamma(-\frac{2 j + 2}{k+2})}} \; = \;
\frac{\gamma_{-}}{\gamma_{+}},
\end{equation}
where $\gamma_{\pm}$ is defined in \cite[p. 349]{TK}.
\bigskip

\section{Reconstruction}
\renewcommand{\theequation}{6.\arabic{equation}}
\setcounter{equation}{0}

In section~2, I explained how the chiral subtheory corresponding to a
WZW conformal field theory can be constructed.  The chiral Hilbert
space, which was obtained in this way, is larger than the
direct sum of the representations of the chiral algebra, and similarly
--- as there exists a correspondence between vertex operators and
states --- the chiral vertex operators depend on internal degrees of
freedom.

In this section I want to describe how one can reconstruct the
original conformal field theory from its chiral subtheory. I will show
that there exists some sort of ``identity operator'' (related to the
diagonal theory, i.~e.\ to the theory where ${\cal H}_{l}$ is
conjugate to $\overline{\cal H}_{l}$), the action of which on a chiral
theory reconstructs the corresponding original theory.  At first this
might seem a bit surprising, as naively, the whole theory should be
obtained by putting together the holomorphic and the anti-holomorphic
theory. However, as I have defined the (holomorphic)
chiral subtheory to contain the
zero modes of the anti-holomorphic representation spaces, it already
retains sufficient information about the original theory.

As a rough analogy this is similar to the process of reconstructing a
component of a (multi-component) group from a single point in this
component and the action of the identity component of the group.
\smallskip

To explain this in more detail, recall that for a given affine algebra
there always exists a modular invariant diagonal theory, which is the
unique WZW-theory corresponding to the simply connected group
\cite{GW,FGK1,FGK2,T}. If all representations are self-conjugate, this
theory plays the r\^{o}le of the identity operator. Otherwise, we
consider the theory which is related to the diagonal theory by the
automorphism of the fusion rule algebra, which interchanges a
representation with its conjugate. We denote the (anti-holomorphic)
chiral theory corresponding to this theory as
\begin{equation}
\overline{\cal H}^{0} = \bigoplus_{m} U_{m} \otimes
\overline{\cal H}^{0}_{m},
\end{equation}
where the $U_{m}$ are representations spaces of the corresponding
quantum group $U_{q}(g)$. By construction, the
subspace of lowest energy vectors in $\overline{\cal H}^{0}_{m}$
is conjugate to $U_{m}$ as a representation of $U_{q}(g)$.
\medskip

Consider now a (not necessarily diagonal) WZW-model
corresponding to the same affine algebra $\hat{g}_{k}$. We denote
the Hilbert space of this theory by
\begin{equation}
{\cal H}= \bigoplus_{l} {\cal H}_{l} \otimes \overline{\cal H}_{l}
\end{equation}
and the corresponding (holomorphic) chiral theory by
\begin{equation}
{\cal H}_{chir} = \bigoplus_{l} {\cal H}_{l} \otimes
\overline{W}_{l}.
\end{equation}
We want to reconstruct the original theory ${\cal H}$ from
its chiral subtheory ${\cal H}_{chir}$. To this end
we consider the product space
\begin{eqnarray}
{\displaystyle \widetilde{\cal H} :} &  = &
{\displaystyle {\cal H}_{chir} \otimes \overline{\cal H}^{0}}
\nonumber \\
& = &
{\displaystyle \bigoplus_{l} \bigoplus_{m} \;
{\cal H}_{l} \otimes \overline{W}_{l} \otimes
U_{m}\otimes \overline{\cal H}^{0}_{m},}
\end{eqnarray}
on which there is a natural action of the quantum group
$U_{q}(g)$, given by
\begin{equation}
\label{action}
U_{q}(g) \ni a \mapsto \sum \bbbone\otimes \Delta_{q}^{(1)}(a)
\otimes \Delta_{q}^{(2)}(a) \otimes \bbbone,
\end{equation}
where we have used the same notation as in (\ref{not}).
We can reconstruct the original Hilbert space ${\cal H}$
by restricting the product space $\widetilde{\cal H}$
to the subspace, which is
invariant under the action of the quantum group
$U_{q}(g)$ (\ref{action}). It is easy to see that this space has the
right size.
\smallskip

To reconstruct the original vertex operator corresponding to
$\psi\otimes \overline{\psi} \in {\cal H}_{j}\otimes \overline{\cal
H}_{j}$, we consider the tensor product of vertex operators
\begin{equation}
V^{q}(\psi\otimes\bar{w},z) \otimes V^{q}(w\otimes\overline{\psi},\bar{z}),
\end{equation}
where $\overline{w}\in \overline{W}_{j}$, $w\in U_{r(j)}$ and we
regard $\overline{\psi}$ as an element of $\overline{\cal
H}^{0}_{r(j)}$. Here, $r(j)$ is defined by the condition that
$\overline{\cal H}^{0}_{r(j)}$ is the (unique) representation
isomorphic to $\overline{\cal H}_{j}$. Then $\overline{W}_{j}$ and
$U_{r(j)}$ are conjugate representations of the quantum group.  These
tensor products act naturally on $\widetilde{\cal H}$.  Furthermore,
there exists a unique (up to scalar multiple) linear combination of
these tensor products of vertex operators, which leaves the subspace
${\cal H}$ of $\widetilde{\cal H}$ invariant. (In fact, this linear
combination corresponds to the unique vacuum vector in the tensor
product of the two $U_{q}(g)$-representations.) We reconstruct the
whole vertex operator by restricting the action of this linear
combination to the subspace ${\cal H}$.

This determines the vertex operators up to scalar multiples, which can
be absorbed into the normalisation of the fields. For generic $q$, the
reconstructed theory satisfies the unmodified duality property, is
local and therefore agrees with the original theory. This remains true
in the truncated case. It should be noted, that the construction
preserves the one-to-one correspondence between states and vertex
operators at all stages.

\section{Conclusions}
\renewcommand{\theequation}{7.\arabic{equation}}
\setcounter{equation}{0}

In this paper I have explained how a chiral theory with a proper
Hilbert space formulation can be defined for the WZW conformal
field theory. The Hilbert space of the (holomorphic) chiral theory is
larger than the direct sum of the chiral
representations spaces, the extra degrees of freedom being the
lowest energy states of the anti-holomorphic representation spaces.
I have shown that there is a natural action of the truncated quantum
group $U_{q}(g)$ on these internal degrees of freedom, such that
the truncated quantum group plays the r\^{o}le of an internal symmetry
algebra for the chiral theory (for integer $k$). As a consequence of
the truncation, the truncated quantum group is only quasi
coassociative and the chiral theory possesses only a modified duality
property. The original theory can be recovered from the chiral theory
by some sort of gauging procedure.

The construction of the internal symmetry algebra is explicit and explains
the significance of the quantum group symmetry independent
of any specific construction of the theory. It has furthermore the
virtue --- in contrast to some earlier attempts --- of fixing the
$q$-parameter in terms of the level of $\hat{g}$.
\medskip

It should be possible to generalise the construction to arbitrary
conformal field theories. In the general case the r\^{o}le of the
internal parameter spaces is played by the ``special subspaces'',
recently introduced by Nahm \cite{Nahm}. (For the WZW-models
these spaces are just the lowest energy states.) This should
provide a constructive way of finding the internal symmetry of a
conformal field theory; details remain to be worked out.
\pagebreak

\noindent {\bf Acknowledgements}

I am grateful to my PhD supervisor Peter Goddard for many helpful
remarks. I also acknowledge useful discussions with M. Chu, H. Kausch,
S. Majid and G.M.T. Watts.

\noindent I am grateful to Pembroke College, Cambridge, for a
research studentship.

\end{document}